\documentstyle[11pt,moriond,epsfig]{article}
\def\Journal#1#2#3#4{{#1} {\bf #2}, #3 (#4)}

\def\PRL{\em Phys. Rev. Lett.}

\def\be{\begin{equation}}
\def\ee{\end{equation}}
\def\bea{\begin{eqnarray}}
\def\eea{\end{eqnarray}}

\begin{document}

\title{SPIN ACCUMULATION IN FS JUNCTIONS}

\author{R. M\'ELIN AND O. BOURGEOIS}

\address{Centre de Recherches sur les Tr\`es Basses Temp\'eratures,
CRTBT,\\
CNRS, BP166X, 38042 Grenoble Cedex, France\\
Laboratoire conventionn\'e avec l'Universit\'e Joseph Fourier}

\maketitle\abstracts{We consider Andreev
reflection in ferromagnet-superconductor
junctions and extend the treatment by de Jong and Beenakker
to incorporate the effect of a repulsive potential at the
interface. We show the existence of a spin accumulation
at a finite voltage in this model. 
}

\section{Introduction}

The interplay between spin polarization and superconductivity
has recently gained a renewed interest because of the
nowadays possibility to study Andreev reflection~\cite{Andreev}
in small junctions involving ferromagnetic
metals in ballistic~\cite{Soulen,Upahyay}
as well as diffusive systems~\cite{Giroud}.
de Jong and Beenakker~\cite{deJong} 
studied multichannel effects in Andreev reflection 
in ferromagnet-superconductor junctions
with a transparent interface. 
Fal'ko {\sl at al.}~\cite{Falko}
and Jemeda {\sl et al.}~\cite{vanWess}
have recently shown the existence of
a spin accumulation in a diffusive
ferromagnet-superconductor interface model.
We consider here ballistic ferromagnet-superconductor
interfaces and extend the de Jong and Beenakker
treatment~\cite{deJong} to incorporate a
finite repulsive interfacial potential.
We show the existence of a  spin accumulation
also in this model.

\section{One channel FS junction model}
We first consider a one channel model describing a contact
between a ferromagnet and a superconductor. Both are thermally
equilibrated reservoirs.
Their energy distribution is therefore
given by the equilibrium Fermi-Dirac distribution.
We consider a Stoner
ferromagnet, modeled by non interacting electrons
in the presence of an exchange field $h(x) = h \theta(-x)$.
The polarization enters this model via
the ratio $h/\epsilon_F$ of the exchange field
$h$ over the Fermi energy $\epsilon_F$.
We also consider the existence of an interfacial
repulsive potential $H \delta(x)$ at the interface.
We introduce the dimensionless barrier strength
$Z=H / \hbar v_F$~\cite{BTK}. For a metallic junction
$Z=0$ and $Z= + \infty$ for a tunnel junction.
The superconductor is a BCS superconductor
with a gap $\Delta(x) = \Delta
\theta(x)$.
The scattering problem will be solved following
Blonder, Thinkham and Klapwick (BTK)~\cite{BTK}.
We consider an electron with an energy $E$
incoming on the junction. Since this electron
also has a spin energy term, its wave vector
depends on spin polarization:
$$
\hbar k_F^{(\pm)} = \sqrt{2 m}\sqrt{E + \mu \pm h}
.
$$
Whereas de Jong and Beenakker considered 
multichannel effects with $Z=0$~\cite{deJong},
we focus here
on the case of a finite value of $Z$, similarly to
\v{Z}uti\'c and Valls~\cite{Zutic}.
Let us consider an incoming electron with a spin up
and match the wave functions on the left and the
right of the F-S contact at a constant energy $E$.
The left-side wave function is
$$
\psi_L(x) = \left( \begin{array}{c}1\\0\end{array} \right)
e^{i k_F^\uparrow x} + a^{e \uparrow}(Z^\uparrow,Z^\downarrow,E) \left(
\begin{array}{c} 0\\1 \end{array} \right) e^{i k_F^\downarrow x} 
+ b^{e \uparrow}(Z^\uparrow,Z^\downarrow,E)
 \left( \begin{array}{c} 1\\ 0 \end{array} \right)
e^{-i k_F^\uparrow x}
$$
while the right-side wave function is
$$
\psi_R(x) = c \left( \begin{array}{c} u_0 \\ v_0
\end{array} \right) e^{i k_F x} +
d \left( \begin{array}{c} v_0 \\ u_0
\end{array} \right)e^{-i k_F x}
,
$$
where the superscript ``$e \uparrow$'' refers to an incoming
electron with a spin up and $u_0$ and $v_0$ denote
the usual BCS coherence factors.
The Andreev reflection amplitude
$a$, the backscattering amplitude $b$ and the transmission
amplitudes $c$ and $d$ are found by matching the
wave function and its derivative according to
\begin{eqnarray}
&& \psi_R(0) = \psi_L(0)\\
&& \frac{\hbar^2}{2 m} \left( \frac{\partial \psi(0^+)}
{\partial x} - \frac{\partial \psi(0^-)}
{\partial x} \right) = H \psi(0)
.
\end{eqnarray}
This leads to
\begin{eqnarray}
\label{eqmatch1}
1+b^{e \uparrow} &=& c u_0 + d v_0 = \frac{i}{2 Z}(c u_0 - d v_0)
- \frac{i}{2 Z^\uparrow}(1-b^{e \uparrow})\\
a^{e \uparrow} &=& c v_0 + dd u_0 = \frac{i}{2 Z}(c v_0 - d u_0)
- \frac{i}{2 Z^\downarrow} a^{e \uparrow}
\label{eqmatch2}
,
\end{eqnarray}
with a different barrier $Z^\uparrow=m H / \hbar^2 k_F^\uparrow$
and $Z^\downarrow = m H / \hbar^2 k_F^\downarrow$ for spin up
and spin down electrons.
Solving for the Andreev reflection coefficient, we find
\begin{equation}
\label{eq:aeup}
a^{e \uparrow}(Z^\uparrow,Z^\downarrow,E) = \frac{4 u_0 v_0} {\left\{ 
(u_0^2 - v_0^2 ) \left( \frac{Z^{\uparrow}}{Z} + 4 Z Z^{\uparrow}
+ \frac{Z}{Z^{\downarrow}} \right)
+  1 + \frac{Z^{\uparrow}}{Z^{\downarrow}}
+ 2 i (u_0^2 - v_0^2) Z
\left( \frac{Z^{\uparrow}}{Z^{\downarrow}} - 1 \right) \right\}}
.
\end{equation}
This expression coincides with the BTK Andreev
amplitude~\cite{BTK} if $Z^\uparrow = Z^\downarrow = Z$.

Let us now consider an incoming hole in the spin down
band. This will be Andreev into an electron with a spin up,
therefore contributing to the current of left-moving
electrons with a spin up. The left-side wave
function is
$$
\psi_L(x) = \left( \begin{array}{c}0\\1\end{array} \right)
e^{- i k_F^\downarrow x} + a^{h \downarrow}(Z^\uparrow,Z^\downarrow,E) \left(
\begin{array}{c} 1\\0 \end{array} \right) e^{- i k_F^\uparrow x} 
+ b^{h \downarrow}(Z^\uparrow,Z^\downarrow,E)
\left( \begin{array}{c} 0\\ 1 \end{array} \right)
e^{i k_F^\downarrow x}
$$
and the right-side wave function is
$$
\psi_R(x) = c \left( \begin{array}{c} v_0 \\ u_0
\end{array} \right) e^{-i k_F x} +
d \left( \begin{array}{c} u_0 \\ v_0
\end{array} \right) e^{i k_F x} 
.
$$
The corresponding matching equations are
\begin{eqnarray}
\label{eq:ahd1}
a^{h \downarrow}(Z^\uparrow,Z^\downarrow,E) &=& c v_0 + d u_0 =
-\frac{i}{2 Z}(c v_0 - d u_0)
+ \frac{i}{2 Z^\uparrow} a^{h \downarrow}(Z^\uparrow,Z^\downarrow,E)\\
1+b^{h \downarrow}(Z^\uparrow,Z^\downarrow,E) &=& c u_0 + d v_0 =
- \frac{i}{2 Z}(c u_0 - d v_0)
+ \frac{i}{2 Z^\downarrow}(1-b^{h \downarrow}(Z^\uparrow,Z^\downarrow,E))
.
\label{eq:ahd2}
\end{eqnarray}
Comparing Eqs.~\ref{eq:ahd1} and~\ref{eq:ahd2} with
Eqs.~\ref{eqmatch1} and~\ref{eqmatch2},
the Andreev coefficient $a^{h \downarrow}(Z^\uparrow,Z^\downarrow,E)$
of an incoming hole with a spin down
is deduced from Eq.~\ref{eq:aeup} with
the prescription
$Z \rightarrow - Z$, $Z^\uparrow \rightarrow - Z^\downarrow$
and $Z^\downarrow \rightarrow - Z^\uparrow$. Since the incoming
and outcoming wave vectors have been interchanged, we
interchange $Z^{\uparrow}$ and $Z^{\downarrow}$, and we
change sign because we are using hole instead
of particle wave functions.
We obtain
$
a^{h \downarrow}(Z^\uparrow,Z^\downarrow,E)
\equiv (k^\downarrow / k^\uparrow) a^{e \uparrow}(Z^\uparrow,Z^\downarrow,E)
$.
This identity is expected because of particle-hole
conjugation.

Now an incoming electron with a spin down has an
Andreev reflection amplitude
$$
a^{e \downarrow}(Z^\uparrow, Z^\downarrow,E) = a^{e \uparrow}(Z^\downarrow,Z^\uparrow,E)
,
$$
which is not related in a simple way to
$a^{e \uparrow}(Z^\uparrow,Z^\downarrow,E)$ below the superconducting
gap because of the phase shift term in Eq.~\ref{eq:aeup}.
A final set of identities is given by the amplitude current
conservation, already discussed by \v{Z}uti\'c and Valls~\cite{Zutic}:
\begin{equation}
\label{eq:current}
k^\uparrow = k^\downarrow |a^{e \uparrow}(Z^\uparrow,Z^\downarrow,E)|^2
+ k^\uparrow |b^{e \uparrow}(Z^\uparrow,Z^\downarrow,E)|^2
.
\end{equation}
Let us now discuss the consequences for the zero temperature
current below the superconducting gap. We find
\begin{eqnarray}
I^{\uparrow}(V)
&=& \int dE f_0(E-eV)
- \int dE \left( \frac{k^\uparrow}{k^\downarrow}\right)
 |a^{h \downarrow}(Z^\uparrow,Z^\downarrow,E)|^2
(1-f_0(-E-eV)) \\
&-& \int dE |b^{e \uparrow}(Z^\uparrow,Z^\downarrow,E)|^2
f_0(E-eV)\\
&=&  \int dE \left( \frac{k^\downarrow}{k^\uparrow} \right)
|a^{e \uparrow}(Z^\uparrow,Z^\downarrow,E)|^2 (f_0(E-eV)-f_0(E+eV))
,
\end{eqnarray}
where $f_0=\theta(-E)$ denotes the zero temperature Fermi
distribution. 
Similarly,
$$
I^{\downarrow}(V)
= \int dE \left( \frac{k^\uparrow}{k^\downarrow} \right)
|a^{e \downarrow}(Z^\uparrow,Z^\downarrow,E)|^2 (f_0(E-eV)-f_0(E+eV))
.
$$
The current of spin down electrons is not equal to
the current of spin up electrons because of the phase shift
term in the expression Eq.~\ref{eq:aeup} of the Andreev reflection
coefficient,
therefore leading
to a spin accumulation at finite voltages. If we consider
the limit of a transparent interface
($Z=0$, $Z^\uparrow=0$ and $Z^\downarrow=0$ while $Z^\uparrow/Z$ and
$Z^\downarrow/Z$ remain finite) we find 
$a^{e \downarrow}(Z^\uparrow,Z^\downarrow,E)= (k^\downarrow / k^\uparrow)
a^{e \uparrow}(Z^\uparrow,Z^\downarrow,E)$,
showing that the current is not spin polarized 
if the interface is transparent.

\section{Spin accumulation}
\begin{figure}
\centerline{\psfig{figure=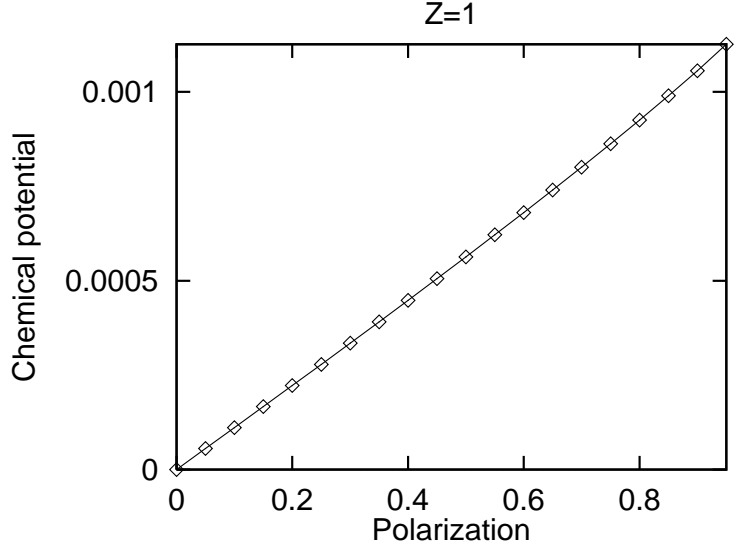,height=3.2in}}
\caption{Variation of the chemical potential 
$\delta \mu = \delta \mu^\uparrow(0) = - \delta \mu^\downarrow(0)$
as a  function of spin polarization for $Z=1$. The chemical
potential is measured in units of the superconducting
gap $\Delta=1$. The voltage is $V=0.1$.
\label{fig}
}
\end{figure}

Since a spin polarized current cannot enter below the
superconducting gap, we should consider a mechanism
restoring a spin unpolarized current at finite voltages.
We consider spin flip
processes in the ferromagnet giving rise to
a different chemical potential for spin up and spin down carriers
at the F-S interface:
$\delta \mu^{\uparrow}=- \delta \mu^\downarrow=\delta \mu$. We will solve for
$\delta \mu$ and show that $\delta \mu >0$.
This existence of a spin accumulation
was put forward by
Fal'ko {\sl et al.}~\cite{Falko} and
Jedema {\sl et al.}~\cite{vanWess}
in the diffusive regime and we show the existence of a 
spin accumulation also in the ballistic regime.
The currents of spin up and spin down electrons are
\begin{eqnarray}
\label{eq:Iupdmu}
I^{\uparrow}(\delta \mu,V) &=&
\int dE \left(\frac{k^\downarrow}{k^\uparrow} \right)
|a^{e \uparrow}(Z^\uparrow,Z^\downarrow,E)|^2 \left[
f_0(E-eV- \delta \mu) - f_0(E+e V + \delta \mu)
\right]\\
I^{\downarrow}(\delta \mu,V) &=&
\int dE \left(\frac{k^\uparrow}{k^\downarrow} \right)
|a^{e \downarrow}(Z^\uparrow,Z^\downarrow,E)|^2 \left[
f_0(E-eV+ \delta \mu) - f_0(E+e V - \delta \mu)
\right]
\label{eq:Idodmu}
,
\end{eqnarray}
Physically, an outcoming electron
with a spin up can be produced by backscattering
or Andreev reflection. Backscattering is controlled
by the chemical potential $+ \delta \mu$
of the spin up electrons whereas Andreev
reflection is controlled by the chemical potential
$- \delta \mu$ of the spin down electrons.
The chemical potential $\delta \mu$ is then determined
by imposing a zero spin current
entering below the superconducting gap:
$I^{\uparrow}(\delta \mu,V)=I^{\downarrow}(\delta \mu,V)$.
We solved this equation in the
small-$\mu$, small-$V$ regime, in which case
the integrals in Eqs.~\ref{eq:Iupdmu}
and~\ref{eq:Idodmu} can be expanded in
the small parameter $eV+\delta \mu$.
We note $g^\uparrow(E)=(k^\downarrow/k^\uparrow) |a^{e \uparrow}(Z^\uparrow,Z^\downarrow,E)|^2$
and make the following approximation for
the integral Eq.~\ref{eq:Iupdmu} of the current:
\begin{equation}
\label{eq:approx}
\int_0^{eV + \delta \mu} g^\uparrow(E) dE \rightarrow \frac{1}{2}
\left( g^\uparrow(0) + g^\uparrow(eV + \delta \mu) \right)
(eV + \delta \mu)
.
\end{equation}
We obtain
\begin{eqnarray}
I^\uparrow &=& \left[ g^\uparrow(0) + g^\uparrow(eV + \delta \mu) \right]
(eV + \delta \mu)\\
I^\downarrow &=& \left[ g^\downarrow(0) + g^\downarrow(eV - \delta \mu) \right]
(eV - \delta \mu)
.
\end{eqnarray}
If only the first terms $g^\uparrow(0)=g^\downarrow(0)$ had
been included, one would have $\delta \mu=0$
whereas we find
a finite and positive $\delta \mu$ when the second
terms are included. If $V$ is finite and $\delta \mu=0$, a  spin
current is present
($I^\uparrow < I^\downarrow$) and the value of
$\delta \mu$ is determined by
equating the spin up and spin down currents.
The variations of $\delta \mu$ as a function of
the spin polarization $P$ are shown on Fig.~\ref{fig}.

We have approximated the
integral Eq.~\ref{eq:Iupdmu} according to Eq.~\ref{eq:approx}.
This approximation
does not modify the qualitative behavior because
$g^\uparrow(E)$ and $g^\downarrow(E)$ are increasing
with energy. Even without this
approximation, the spin up current increases
upon increasing $\delta \mu$, and the
spin down current decreases, therefore leading
to a finite $\delta \mu$ at finite voltages.

A finite value of $\delta \mu$ at the interface 
can be generated by
spin flip processes in the ferromagnet.
If we note $f_R^\sigma(E,x)=f_T(E-\mu^\sigma(x))$ the
distribution of right moving electrons in the
ferromagnet ($x<0$) the semi classical Boltzmann equation
with a spin flip collision term is
\begin{eqnarray}
\frac{\partial}{\partial x} f_R^\uparrow(E,x)
&=& - \frac{1}{l_s}
\left(f_R^\uparrow(E,x) - f_R^\downarrow(E,x) \right)\\
\frac{\partial}{\partial x} f_R^\downarrow(E,x)
&=& - \frac{1}{l_s}
\left(f_R^\downarrow(E,x) - f_R^\uparrow(E,x) \right)
,
\end{eqnarray}
with $l_s$ the spin flip length and the boundary
condition $\delta \mu^\uparrow(0)=
- \delta \mu^\downarrow(0)=-\delta \mu$.
From what we deduce
$
\delta \mu^{\pm}(x) = \mp e^{2 x / l_s} \delta \mu
$.

\section{Multichannel model}
Let us now consider multichannel effects
discussed by de Jong and Beenakker in the
case of a metallic contact~\cite{deJong}. We consider
a number $N^\downarrow \sim (k^\downarrow)^2$ of spin down channels,
and a number $N^\uparrow \sim (k^\uparrow)^2 > N^\downarrow$ of
spin up channels. We consider all the
spin up and spin down channels to have the same Fermi
wave vector $k_F^\uparrow$ and $k_F^\downarrow$.
The transport equations below the superconducting gap are
$f_R^\sigma(E) = N^\sigma f_0(E-e V)$, and
\begin{eqnarray}
\nonumber
f_L^\uparrow (E) &=& N^\downarrow \left\{ 
\left( {k^\uparrow \over k^\downarrow}\right)
|a^{h \downarrow}(Z^\uparrow,Z^\downarrow,E)|^2
[1-f_0(-E-eV)]
+ |b^{e \uparrow}(Z^\uparrow,Z^\downarrow,E)|^2 f_0(E-eV)\right\}\\
\nonumber
& & + (N^\uparrow - N^\downarrow) f_0(E-eV)\\
f_L^\downarrow(E) &=& N^\downarrow \left\{ 
\left( {k^\downarrow \over k^\uparrow} \right)
|a^{h \uparrow}(Z^\uparrow,Z^\downarrow,E)|^2 [1-f_0(-E-eV)] 
+ |b^{e \downarrow}(Z^\uparrow,Z^\downarrow,E)|^2 f_0(E-eV) \right\}
,
\nonumber
\end{eqnarray}
the last term arising from the fraction
$1-N^\downarrow/N^\uparrow$ of the incoming electrons with a spin up
that are not Andreev reflected and are fully backscattered.
The zero temperature current of spin up and spin down electrons
is
\begin{eqnarray}
\label{eq:I-T0-1}
I^{\uparrow} &=&  N^\downarrow
\int dE \left( \frac{k^\uparrow}{k^{\downarrow}} \right)
|a^{e \uparrow}(Z^\uparrow,Z^\downarrow,E)|^2
\left( f_0(E-eV) - f_0(E+eV) \right) \\
I^{\downarrow} &=&  N^\downarrow
\int dE \left( \frac{k^\downarrow}{k^{\uparrow}} \right)
|a^{e \downarrow}(Z^\uparrow,Z^\downarrow,E)|^2
\left( f_0(E-eV) - f_0(E+eV) \right)
\label{eq:I-T0-2}
.
\end{eqnarray}
This expression is equal to the single channel current
times the number of spin down channels. A similar mechanism
of spin accumulation also exists in this
model, the spin accumulation being independent
on the number of channels. However $\delta \mu$
is small even though finite
and therefore should not give a significant
contribution to the junction resistance.

\section{Conclusions}
We have analyzed transport in resistive F-S junctions
and shown the existence of
a small, even though finite spin accumulation at finite voltages.
A spin accumulation was obtained by
Fal'ko {\sl et al.}~\cite{Falko}
and Jedema {\sl et al.}~\cite{vanWess}
in contacts between a diffusive ferromagnet and a
superconductor.
Whereas in their model spin accumulation originates
from a different spin conductivity of spin up
and spin down electrons, spin accumulation originates
in our model from a different interfacial
barrier of spin up and spin down electrons.

\section*{Acknowledgments}
The authors acknowledge fruitful discussions with 
D. Feinberg, F. Hekking. The authors thank
V.I. Volkov and
B.J. van Wees for drawing our attention to their
works~\cite{Falko,vanWess} on diffusive
ferromagnet-superconductor junctions.


\section*{References}
\begin{thebibliography}{99}

\bibitem{Andreev}A. Andreev, \Journal{JETP}{19}{1228}{1964}.

\bibitem{Soulen}R.J. Soulen Jr., J.M. Byers, M.S. Osofsky,
B. Nadgorny, T. Ambrose, S.F. Cheng, P.R. Broussard,
C.T. Tanaka, J. Nowak, J.S. Moodera, A. Barry, and
J.M.D. Coey, \Journal{\em Science}{282}{85}{1998}.

\bibitem{Upahyay}S.K. Upahyay, A.P. Palamisami. R.N. Louie, and
R.A. Buhrman, \Journal{\PRL}{81}{3247}{1998}.

\bibitem{Giroud} M. Giroud, H. Courtois, K. Hasselbach, D.
Mailly, and B. Pannetier, Phys. Rev. B {\bf 58}, R11872 (1998).

\bibitem{deJong}M. de Jong and C. Beenakker,
\Journal{\PRL}{74}{1657}{1995}.

\bibitem{Falko} V.I. Fal'ko, C.J. Lambert, 
and A.F. Volkov, JETP Letters {\bf 69},
532 (1999).

\bibitem{vanWess} F.J. Jedema, B.J. van Wees,
B.H. Hoving, A.T. Filip, and T.M. Klapwick,
Report No cond-mat/9901323.

\bibitem{BTK}G. Blonder, M. Tinkham, and T.
Klapwijk, \Journal{{\em Phys. Rev.} B}{25}{4515}{1982}.

\bibitem{Zutic} I. \v{Z}uti\'c and O.T. Valls,
Report No cond-mat/9902080.


\end{thebibliography}
\end{document}